\documentstyle[11pt]{article}

\def\m{\mu}

\def\be{\begin{equation}}
\def\ee{\end{equation}}

\begin{document}

\begin{flushright}
BRX TH-467
\end{flushright}

\begin{center} {\Large\bf Closed Form Effective Conformal Anomaly
Actions in D$\geq$4}

S. Deser\\
Department of
Physics, Brandeis University, Waltham, MA 02454, USA
\end{center}

\begin{abstract}I present, in any D$\geq$4, closed-form type B
conformal anomaly effective actions incorporating the logarithmic
scaling cutoff dependence that generates these anomalies. Their
construction is based on a novel class of Weyl-invariant tensor
operators. The only known type A actions in D$\geq$4 are
extensions of the Polyakov integral in D=2; despite contrary
appearances, we show that their nonlocality does not conflict with
general anomaly requirements. They are, however, physically
unsatisfactory, prompting a brief attempt at better versions.

\end{abstract}


\section{Introduction}

Conformal (Weyl) anomalies reflect the loss of classical scale
invariance caused by unavoidable regularization of conformally
invariant matter closed loops. Their properties, in all
dimensions, are by now well understood. In particular \cite{sdas}
there are two, explicitly known, types that can be conveniently
expressed in terms of an external metric $g_{\mu\nu}$ coupled to
the matter. These local gravitational scalar densities ${\cal
A}(g_{\mu\nu})$ differ in their separate IR/UV origins and in
their behavior under Weyl variations.  Both can be represented as
responses of (nonunique) nonlocal effective actions
$I[g_{\mu\nu}]$ under metric conformal variations
\begin{equation}
\delta g_{\mu\nu} = 2 \phi (x) g_{\mu\nu} \: .
\end{equation}
The built-in integrability condition on variations of these
anomalies
 ${\cal A}(x) \equiv \delta I [g_{\mu\nu}
]/\delta \phi (x)$,
\begin{equation}
\delta {\cal A}(x)/\delta \phi (x^\prime )= \delta^2 I/\delta\phi
(x^\prime ) \delta\phi (x) = \delta {\cal A} (x^\prime ) /
\delta\phi (x) \; ,
\end{equation}
serves as a useful check on candidate ${\cal A}$'s and on allowed
forms of the type A actions; type B anomalies, are necessarily
Weyl-invariant, satisfying (2) trivially.

The origin of the anomalies in closed loop graphs imposes
constraints on the actions' dimensionality and nonlocality. These
seem to clash with the only known closed form actions, essentially
the obvious D$\geq$4 generalizations of the D=2 Polyakov action in
both cases (type B starts at D=4).  On the other hand, since
effective actions are not unique -- nonlocal Weyl invariant
dimensionless gravitational functionals will be exhibited -- some
choices will be better behaved physically than others, reflecting
more accurately the underlying loop properties or being obtainable
through integrating out a compensating field in an action that is
physically acceptable, in particular ghost-free.

I will present here new complete closed form type B actions that
correctly reflect their cutoff dependence and origins.  Their
construction is based on new tensor differential operators
(generalizing existing scalar ones) that are  conformal invariant
when acting on Weyl-like tensors. For the existing type A actions,
we will resolve the paradox that their explicit $\Box^{-2}$
nonlocalities violate the single pole $(\Box^{-1} )$ behavior at
lowest order about flat space required by dimensionality and
general anomaly analysis. Quite apart from the above problems,
however, they have long been known \cite{osborn,sdas}, to be
unsatisfactory in {\it e.g.}, long-distance behavior, failing to
correctly represent the underlying stress tensor correlators.
While I have not succeeded in constructing more suitable actions
beyond the lowest order one given in \cite{sdas}, some remarks on
this open problem are appended.


\section{Type B}

The type B anomalies ${\cal A}_B$ have two hallmarks: they arise
from the UV behavior of the underlying matter loops, with
consequent logarithmic cutoff dependence, and are themselves Weyl
invariant.  They only start at D=4, being in fact the first
anomalies discovered there \cite{sdmd}; the unique local D=4
conformal invariant is the square of the Weyl tensor,
\begin{equation}
{\cal A}_B \equiv \sqrt{-g} \: tr \: C^2 \; ,\;\;\; %
\delta {\cal A}_B (x)/\delta \phi (x^\prime ) \equiv 0 \; .
\end{equation}%
The number of independent ${\cal A}_B$ rises rapidly with
dimension; for example there are  \cite{armenians} three varieties
at D=6: two independent index traces of $\sqrt{-g} \: C^3$ and a
third of the schematic form $\sqrt{-g}\: tr C (\Box + R) C$. An
effective D=4 action that reflects the required logarithmic
behavior was already introduced in \cite{sdmd} at lowest, cubic,
order in an expansion about flat space, $h_{\mu\nu} \equiv
g_{\mu\nu} - \eta_{\mu\nu}$:
\begin{equation}
I^4_B \approx \int d^4x \, tr \, C \ln (\Box/\lambda^2 ) C + {\cal
O} (h^4) \; ,
\end{equation}%
each $C$ and $\Box$ being effectively ${\cal O} (h)$, since the
quadratic part is Weyl invariant. While not strictly correct, this
approximation reproduced some of the desired scaling
characteristics, including the logarithmic dependence on the
cutoff $\lambda$ of the closed loops. What we really want of
course is to retain that behavior, but with $\Box$ replaced by an
argument $\tilde{\Delta}$ that produces the proper variation
$\delta \ln \tilde{\Delta} = \phi$ to all orders; $\tilde{\Delta}$
must be a scalar (covariance forbids densities from being
arguments of logs) and of dimension 4, to bring in a
$\lambda^{-4}$. The problem of obtaining a physical effective
action $I_B$ thus reduces to finding dimension 4 operators
$\tilde{\Delta}_4$ that, at least when acting on a specific
tensorial class $Z$ such as scalar or 4-tensors, are themselves
Weyl invariant, $(\delta \tilde{\Delta}_4) Z = 0$.  One such
operator has long been know and indeed underlies the type A
construction to be discussed in Sec. 3: the self-adjoint Paneitz
(scalar density) operator \cite{paneitz} acting on scalars,
\begin{equation}
\Delta_P = \sqrt{-g} [\Box^2 + 2 D_\mu (R^{\mu\nu} -
\textstyle{\frac{1}{3}} \, g^{\mu\nu} R) D_\nu  ] \;.
\end{equation}%
It is the unique D=4 generalization of the D=2 invariant $\Delta_2
\equiv \sqrt{-g} \: \Box$; its extra terms complete the merely
constant-scale invariant $\sqrt{-g} \: \Box^2$. Unfortunately,
because it acts on scalars, $\Delta_p$ is useless here:
\begin{equation}
\tilde{I}^4_B \sim \int d^4x \, \sqrt{-g}\: \ln (\Delta_P /
\sqrt{-g} \:
\lambda^4) tr\, C^2 %
\end{equation}%
is a total divergence, with vanishing variation.

From the above lesson, it is clear that we must abandon invariant
$\Delta$'s acting on scalars and instead seek one that begins at
D=4 and acts invariantly on 4-tensors $T$, $\delta
(\tilde{\Delta})T =0$. More specifically, it suffices that
$\tilde{\Delta}$ be invariant when acting on the Weyl tensor, for
concreteness in its Weyl invariant $C^\m_{~\alpha\beta\gamma}$
index configuration, and to reproduce the latter's tensorial rank
and algebraic properties:
\begin{equation}
\tilde{C}^\m~_{\nu\alpha\beta} \equiv (\tilde{\Delta}
C)^\mu_{\nu\alpha\beta} \equiv
\tilde{\Delta}^{\mu\nu{^\prime}\alpha{^\prime}\beta{^\prime}}_{\mu{^\prime}\nu\alpha\beta}
C^{\mu{^\prime}}_{~\nu{^\prime}\alpha{^\prime}\beta{^\prime}}\;.
\end{equation}%
Preserving constant scale invariance already requires
$\tilde{\Delta}$ (like $\Delta_p$) to be a 4$^{\rm th}$ derivative
tensor density; so if $\tilde{\Delta}$ obeys $\delta
(\tilde{\Delta}) C=0$, it will follow that
\begin{equation}
\delta\tilde{C} = 0 \; , \;\;\; \delta (\tilde{\Delta} \:
\frac{1}{\sqrt{-g}} \: \tilde{C} ) = 0 \; , \;\;\;
\delta[\tilde{\Delta} \frac{1}{\sqrt{-g}} \; \tilde{\Delta} \; ...
\; \frac{1}{\sqrt{-g}} \: \tilde{\Delta} C)]=0 \; ,
\end{equation}%
where the intermediate $1/\sqrt{-g}$ factors must be included to
keep subsequent $\tilde{\Delta}$ acting on tensors rather than on
densities.  The underlying physics clearly demands that such a
$\tilde{\Delta}$ exist, and it has now indeed been found
\cite{branson}; while its form is unfamiliar ({\it e.g.}, it has
no $\Box^2$ part at all), it is only necessary for our purposes to
know that it exists since its only role is to allow for the
presence of the ``compensator field" ln$\sqrt{-g}$. In terms of
$\tilde{\Delta}_4$ the desired action is simple:
\begin{equation}
I^4_B = - \textstyle{\frac{1}{4}}\int  d^4x \sqrt{-g}\:
C_\mu^{~\nu\alpha\beta} [\ln (\tilde{\Delta}_4/\lambda^4 \sqrt{-g}
\: ) C]^\mu_{\nu\alpha\beta} \; .
\end{equation}%
The only non-vanishing variation of (9) stems entirely from
ln$\sqrt{-g}$,
$$
\delta I^4_B = \textstyle{\frac{1}{4}}\int   d^4x \sqrt{-g}\, C^2
\delta \ln \sqrt{-g} = \int   d^4x (\sqrt{-g} C^2 ) \delta
\phi \; , \eqno{\rm (10a)} $$%
all the rest of (9), including the left factor $(\sqrt{-g} \,
g^{..} g^{..}C^. ...)$, being manifestly invariant.  In more
detail, since it is the density $\tilde{\Delta}_4$ that is Weyl
invariant, any power in the log's expansion, $(\frac{1}{\sqrt{-g}}
\, \tilde{\Delta} ... \frac{1}{\sqrt{-g}} \tilde{\Delta}) C$
correctly avoids having $\tilde{\Delta}_4$ act on densities and
only the ``outer" $1/\sqrt{-g}$  factor contributes in each term.
Hence we may indeed conclude from (10a) that
$$
\delta I^4_B /\delta\phi = \sqrt{-g} \, C^2 \; . \eqno{\rm (10b)}%
$$%
The various  possible ${\cal A}_B(x)$ in higher dimension will
similarly be expressible in terms of the corresponding
$\tilde{\Delta}_D$, which are also sure to exist:%
\renewcommand{\theequation}{\arabic{equation}}
\setcounter{equation}{10}
\begin{equation}
I^d_B   \sim   \int   d^dx  \sqrt{-g}\, Z_\mu^{\,\nu\alpha\beta} [
\ln  (\tilde{\Delta}_D/\lambda^D
\sqrt{-g}\: ) C]^\mu_{\nu\alpha\beta}  %
\end{equation}
where $Z$ is the ``rest" of the local invariant in question, {\it
e.g.}, $Z \sim (CC)$ or $C(\Box +R)$ in D=6 and similarly for
D$>$6.

I close this section with an object lesson on ambiguities in
effective actions; it is an apt introduction to the type A
problem, being modeled on the only closed form action known there
and being even more unphysical for type B (because it totally
violates the logarithmic dependences) than for type A.\  It is
based on the fact that (as explained below) the quantity
$(\bar{\cal E}_4 \Delta^{-1}_p)$, where $\bar{\cal E}_4$ is
essentially the D=4 Euler invariant, Weyl transforms as a
compensator field (21).\ Consequently~\cite{riegert},
\begin{equation}
^*\!I^4_B = \int d^4x \: \bar{\cal E}_4 \Delta^{-1}_p (\sqrt{-g}\:
C^2 )\; , \;\;\; \delta \: ^*\!I^4_B / \delta \phi = \sqrt{-g} \:
C^2 \; .
\end{equation}%
Note the complete contrast between the actions (9) and (12), even
though both succeed in the limited requirement of correctly
yielding ${\cal A}_B$ under Weyl variation.


\section{Type A}

To understand this family, it is useful to review D=2, where type
A is the only possible anomaly. By power counting, the anomaly
${\cal A}_2(x)$ must have dimension 2; the only local
diffeo-invariant is the Euler density ${\cal E}_2 (x)$, a total
divergence:
\begin{equation}
{\cal A}_2(x)  =  \sqrt{-g} \, R(x) = \textstyle{\frac{1}{2}} \:
\sqrt{-g} \: \epsilon^{\mu\nu} \epsilon^{\alpha\beta} R_{\mu\nu\,
\alpha\beta} \equiv {\cal E}_2 (x)\;.
\end{equation}%
Unlike type B, this quantity is not Weyl invariant; rather $\delta
{\cal E}_2 (x) \equiv 2 \sqrt{-g} \, \Box \phi$. The indicated
Weyl variation of ${\cal A}_2$ guarantees the integrability
condition:
\begin{equation}
 \delta {\cal E}_2 (x)/\delta\phi (x^\prime
) = 2 \sqrt{-g}\: \Box \delta^2 (x-x^\prime) = \delta {\cal E}_2
(x^\prime )/\delta \phi (x) \; .
\end{equation}
As already noted, the scalar density operator $\Delta_2 \equiv
\sqrt{-g} \: \Box$ is Weyl invariant at D=2, when (and only when)
acting on a scalar
\begin{equation}
\delta\Delta_2 \equiv \delta [\partial_\mu (\sqrt{-g}\,
g^{\mu\nu})\partial_\nu ]=0 \; . %
\end{equation}%
Hence the nonlocal scalar operator
\begin{equation}
\delta ({\cal E}_2 / \Delta_2 ) = 2 \phi (x)%
\end{equation}
transforms like a Weyl compensator field, leading to the Polyakov
\cite{polyakov} construction,%
\begin{equation}
I_2 = \textstyle{\frac{1}{4}} \: \int \: d^2x\;{\cal E}_2
\Delta^{-1}_2 {\cal E}_2  \; , \;\;\;\; \delta I_2/\delta\phi (x)
= {\cal E}_2 (x) \; . %
\end{equation}%
Note that although $\Delta^{-1}_2$ acts on the density ${\cal
E}_2$, its variation vanishes because we must first write $\delta
\Delta^{-1}_2 {\cal E}_2 = - \Delta^{-1}_2 \: \delta (\Delta_2)
(\Delta^{-1}_2 {\cal E}_2)$ and $(\Delta^{-1}_2 \: {\cal E}_2 )$
is a scalar. The pole behavior of the action is clearly $\sim
p^{-2}$, in accord with the power counting of the 2-point closed
loop $\sim (\int d^2p/p^4)R^2_L$ where $R_L \sim (pph)$ are the
linearized scalar curvatures representing external gravitons
coupled to the matter $T_{\mu\nu}$:  The underlying correlator, $<
T^{\mu\nu}{(p)}T^{\alpha\beta}{(-p)}>$, is multiplied by
$h_{\mu\nu} h_{\alpha\beta}$ and the four factors of momentum in
the $<TT>$ numerator convert them to curvatures. However, this
counting is true only to leading order in $h_{\mu\nu}$: while
``dressings" of the curvatures from expanding (17) in powers of
$h$ but keeping the flat space $\Delta^{-1}_2$ do indeed maintain
the $p^{-2}$ overall behavior (as they should diagrammatically
since this is still effectively a 2-point function)
expanding the denominator $\Delta^{-1}_2$,%
\begin{eqnarray}
\Delta^{-1}_2  & \equiv &  [\Box_0 + \partial_\mu {\cal
H}^{\mu\nu} \partial_\nu ]^{-1} =  (1  -   \Box^{-1}_0\partial_\mu
{\cal H}^{\mu\nu} \partial_\nu +  ... ) \Box^{-1}_0, \;\;\;\;
 \Box_0  \equiv  \eta^{\mu\nu} \partial_\mu\partial_\nu \nonumber \\
 {\cal H}^{\mu\nu} & \equiv &  \sqrt{-g} \, g^{\mu\nu} -
\eta^{\mu\nu} = - (h^{\mu\nu}- \textstyle{\frac{1}{2}} \,
\eta^{\mu\nu} h) + ...
\end{eqnarray}
gives rise to increasing powers of $p^{-2}$, in total agreement
with the diagrammatics; a 3-point closed loop generically acquires
another $p^{-2}$ from the extra propagator and so on for the
$n$-point expansion.  Indeed, the folklore that anomalies must
have only a $\Box^{-1}_0$ nonlocality, applies only to their
leading terms. This fact will be essential in D=4.  Despite these
seemingly unpleasant higher poles, the Polyakov action (unlike its
D$>$2 extensions) is perfectly physical as attested by its
derivability through integrating out a physical ghostfree
compensator field's action, $I_2 [\sigma ] = \int d^2x
[\frac{1}{2} \sigma \Delta_2 \sigma + \sigma {\cal E}_2 ]$; it is
also vouched for by being the covariantization of the matter loop
integrals $\int d^2x h <TT> h$, as noted above.

As shown in \cite{sdas}, the D=2 anomaly (13) extends uniquely to
any D=2$n$: the same ``infrared" type is given by the Euler
density at D=2$n$,
\begin{equation}
{\cal A}_{2n} = {\cal E}_{2n} \equiv
\epsilon^{1..2n}\epsilon^{1'..2n'} R_1...R_{...2n{^\prime}} / \sqrt{-g} \; ; %
\end{equation}%
note that ${\cal E}_{2n}$ and hence its variation vanishes
identically in lower D (since the Levi--Civita $\epsilon^{1...2n}$
symbol does). Integrability is always satisfied because ${\cal
E}_{2n}$ varies according to
\begin{equation}
\delta {\cal E}_{2n} (x)/\delta \phi (x^\prime ) = {\cal
G}^{\mu\nu}_{2n} (x) D_\mu D_\nu \delta (x-x^\prime ) = \delta \,
{\cal
E}_{2n}(x^\prime ) / \delta \phi (x) \; .%
\end{equation}%
where ${\cal G}^{\mu\nu}_{2n}$ is an identically conserved tensor
(as it must be, since ${\cal E}_{2n}$ and its variations are total
divergences); it is essentially the ``Einstein tensor" of the
Euler action at dimension 2($n$-1). For concreteness we will work
in D=4, then indicate the generalization to arbitrary D. Here,
${\cal G}^{\mu\nu}_4$ is of course the true Einstein tensor (that
indeed vanishes at D=2) and ${\cal E}_4 \equiv \sqrt{-g} \:
(R^2_{\mu\nu\alpha\beta} - 4 R^2_{\mu\nu} + R^2 )$ is the usual
Gauss--Bonnet combination. It is therefore tempting to follow the
form of the D=2 action (17), in terms of the suitable
generalization of $\Delta_2$, namely the Paneitz operator (5) and
replacing ${\cal E}_2$ by ${\cal E}_4$. This was the proposal of
\cite{riegert}, with the minor modification of using $\bar{\cal
E}_4 \equiv {\cal E}_4 + \frac{2}{3} \, \sqrt{-g} \, \Box R$
rather than ${\cal E}_4$, to achieve the extension of (16), to
\begin{equation}
\delta (\bar{\cal E}_4 / \Delta_P) = \phi \; ,
\end{equation}
and therefore of (17) to
\begin{equation}
I^A_4  =  \int  d^4x \, \bar{\cal E}_4 \Delta^{-1}_p \bar{\cal
E}_4 \; , \;\;\;\;\; \delta I^A_4/\delta\phi  = \bar{\cal E}_4 \;
,
\end{equation}
by the same reasoning as in D=2.  Since $\Box R$ itself derives
from a local (and hence irrelevant, removable) action, $\delta
\frac{1}{18} \int d^4x \,  \sqrt{-g}\:R^2 / \delta\phi =
\frac{2}{3} \sqrt{-g} \, \Box R$, we see that while (22) literally
varies into $\bar{{\cal E}}_4$, it effectively also varies into
${\cal E}_4$.

The representation (22) presents a paradox: $\Delta^{-1}_p$
contains a double ($\Box^{-2}_0$) pole, whose presence is
incompatible with the leading, 3-point, function. Just by momentum
counting around a matter loop with three external curvatures, the
latter's leading ${\cal O} (h^3)$ term has to be $\Box^{-1}_0$,
and not $\Box^{-2}_0$, nonlocal. To understand the conflict in
detail, first simply expand the $\Delta_p$ of (5) in (22):
\begin{equation}
I^A_4 [h^3] =  \int d^4x ({\cal E}_4 + \textstyle{\frac{2}{3}}\,
\Box_0 R) [1-2\Box^{-2}_0
\partial_\mu (R^{\mu\nu} - \textstyle{\frac{1}{3}} \:
\eta^{\mu\nu} \: R )
 \partial_\nu \! + \! ...] \Box^{-2}_0
({\cal E}_4 +  \textstyle{\frac{2}{3}} \Box_0 R ) \; .
\end{equation}
Here curvatures are needed only to their linearized ${\cal O} (h)$
order and derivatives are also flat space ones; all corrections to
those quantities either lead to ${\cal O} (h^4 )$ or are ${\cal O}
(h^3 )$ but harmless, $\sim \Box^{-1}_0$. The same is true of the
unity part of the $\Delta^{-1}_4$ expansion: the quadratic terms
are the local $\int d^4x \, R^2$, the cubics are $\sim \int d^4x
[{\cal E}_4 \Box_0^{-1} R + \Box_1 R \Box^{-1}_0 R ]$ where
$\Box_1$ is the ${\cal O} (h)$ part of $\Box$; they are
single-pole.  Now pass to the correction term which seems to have
a $\Box^{-4}_0$. However, being linear,  it only multiplies the
quadratics $( \Box_0 R) \Box^{-2}_0 \Box_0 R \sim R^2$, so it is
$\Box_0^{-2}$ at worst. Before proceeding further, note two useful
properties \cite{sdas} of our cubic integrals:  first, the
position of $\Box^{-1}_0$ among the three factors is irrelevant;
second, integration by parts rules are very useful, {\it e.g.},
(for any $S$) $\int d^4x \, S \,
\partial_\mu R \partial^\mu R = - \frac{1}{2} \int d^4x \, S
\Box_0 R^2$.  Both are used implicitly below. The dangerous cubic
terms in (23) reduce to the form
\begin{equation}
\int d^4x \, R_{,\mu} \, \Box^{-1}_0  (R^{\mu\nu} -
\textstyle{\frac{1}{3}} \, \eta^{\mu\nu} R)\Box^{-1}_0 \, R_{,\nu}
\;.
\end{equation}%
The pure $R^3$ part, $\sim \int \eta^{\mu\nu}  R_{,\mu}
\Box^{-1}_0 R \Box^{-1}_0 R_{,\nu} = - \frac{1}{2} \int \! R^3
/\Box_0$ is obviously safe.  This leaves the first,
$R^{\mu\nu}$-dependent one,
\begin{equation}
\int  d^4x R_{,\mu}  \Box^{-1}_0  R^{\mu\nu} \Box^{-1}_0 R_{,\nu}
\; ,
\end{equation}
which is certainly $\sim\Box^{-2}_0$ as it stands. Note that there
is no dimensional contradiction: the extra $\Box^{-1}_0$ is
compensated for by the extra $\partial_\mu
\partial_\nu$ in the numerator, but these are not mutable into a
$\Box_0$ by parts integration as long as we write everything in
terms of curvatures alone. This impasse disappears by relaxing the
latter requirement and expressing the Ricci tensor in terms of its
metric definition,
\begin{equation}
2R_{\mu\nu} = \Box_0 h_{\mu\nu} - (\partial^2_{\mu\alpha}
h^{\alpha~}_\nu + (\nu\mu )) + h^\alpha_{\alpha ,{\mu\nu}} \;,
\;\;\;\; R = \Box_0 h^\alpha_\alpha - \partial^2_{\alpha\beta}
h^{\alpha\beta} \; .
\end{equation}%
The $\Box_0 h_{\mu\nu}$ term is manifestly $\sim\Box^{-1}_0$; the
remaining ones also provide an additional $\Box_0$, after
integration by parts.  The result of a simple calculation yields
the equality
\begin{equation}
\int d^4x \, R_{,\mu} R_{,\nu} \Box^{-2}_0 R^{\mu\nu}=
\textstyle{\frac{1}{2}} \int  d^4x  \left[ R_{,\mu}R_{,\nu}
\Box^{-1}_0 h^{\mu\nu} - \textstyle{\frac{1}{4}}\, R^2
h^\alpha_\alpha + \textstyle{\frac{1}{2}} R^2 \Box^{-1}_0 R
\right] \; .
\end{equation}
whose right side, although its first term is irreducible to
``curvatures/$\Box_0$" is of course just as gauge-invariant under
$\delta h_{\mu\nu}=
\partial_\mu \xi_\nu +\partial_\nu \xi_\mu$, $\delta R=0$ as the
left, after using partial integration.  These steps demonstrate
that the leading, $h^3$, terms (23) have only a simple pole, which
is all one can demand of them on dimensional or general anomaly
grounds. Before proceeding to higher order, it should be
emphasized that the $\Box^{-1}_0$ nonlocality is irreducible, and
cannot be removed even by expressing all curvatures in terms of
$h_{\mu\nu}$.  More important, their failure to have $\int R^3
\Box^{-1}_0$ form is not unavoidable, but rather symptomatic of
their physical defects: that form has been achieved \cite{sdas}.

Beyond leading order there will clearly appear higher and higher
poles in the $h_{\mu\nu}$-expansion of $\Delta^{-1}_p$. Indeed, as
explained previously, each successive additional vertex insertion
into the loop diagram involves an extra propagator and so,
generically an (acceptable) extra power of $\Box^{-1}_0$. For
higher D, the $\Delta_{2n}$ will go as $\Box^n$, but again the
leading, $(n+1)$-point function must go as $\int
(d^{2n}p)/(p^2)^{n+1} \sim p^{-2}$, and it will, by similar
considerations as for D=4, with
\begin{equation}
I_A = \int d^{2n}x \: \bar{{\cal E}}_{2n} \Delta^{-1}_{2n} \:
\bar{{\cal E}}_{2n}
\end{equation}
and $\Delta_{2n} \sim \Box^n +.. \: , \; \bar{{\cal E}}_{2n} \sim
{\cal E}_{2n} + ...$ where the additional terms in $\Delta_n$ and
$\bar{{\cal E}}_{2n}$ are of lower/higher derivative order
respectively, and as in D=2,4 with $\delta\bar{\cal E}_{2n} =
\Delta_{2n}\phi$,~ $\delta \Delta_{2n} = 0$.

The above ``rehabilitation" of  (22) in no way improves its
problematic physical behavior. One illustration of its problems is
supplied by the unphysical nature of the compensating field action
that generates it when the field is integrated out:
\begin{equation}
I_A [\sigma ; g_{\mu\nu} ] = \int d^Dx [\textstyle{\frac{1}{2}} \:
\sigma \Delta_D \sigma + \sigma \bar{{\cal E}}_D ]\; , \;\;\;\;
\delta I [\sigma +\phi \, ; \: 2\phi g_{\mu\nu}]/\delta \phi =
\bar{\cal E}_D \; .
\end{equation}%
This means that for D$>$2, the $\sigma$ propagator becomes more
and more ghostlike, with correspondingly worse long-distance
behavior associated to the higher powers $\Box^n$ in $\Delta_D$.
This correlation is unavoidable: since $\sigma$ must be
dimensionless, its kinetic part has to be of $D$-derivative order.
There are of course an infinite number of compensator actions,
just as there are of purely gravitational ones.  An example of the
former whose variation yields ${\cal E}_4$ is \cite{riegert,sdas}
\begin{equation}
I^\prime_A [\sigma ;g_{\mu\nu} ] = \int d^4x [8 \sigma {\cal E}_4
+ \sigma G^{\mu\nu} D_\mu D_\nu \sigma +
\textstyle{\frac{1}{2}}\Box \sigma (\partial_\mu \sigma )^2 +
(\partial_\mu \sigma )^2 (\partial_\nu \sigma )^2]\; ,
\end{equation}%
each succeeding term correcting the residual variation of the
previous ones; there is no kinetic $\sigma$ term at all. Examples
of ambiguities in the gravitational actions are furnished by
polynomials in the dimensionless conformally invariant scalar
building block $X \equiv \sqrt{-g} \: C^2 \Delta^{-1}_p$,
namely %
\begin{equation}
I_{conf} = \sum^\infty_{m=0} \: a_n \int d^4x X^{m+1} \sqrt{-g}\:
C^2 \; .
\end{equation}
Each term begins at order $h^{2(m+2)}$, with corresponding poles
$\sim \Box^{-2(m+1)}$.

What is clearly required at the purely gravitational action level
is a way of finding (perhaps from the expression given in
\cite{sdas}) the covariantization (without distorting its
behavior) of the lowest order $\int d^4x hhh<TTT>$ action dictated
by the actual loop structure.  The goal would be to generalize the
D=2 action only by increasing the number of curvatures in the
numerator, but keeping the denominator of second order, in terms
of new second-derivative tensor operators $\tilde{\Delta}$ (not
necessarily Weyl invariant) that would permit fully covariant
actions of the form
\begin{equation}
\tilde{I}_A = \int d^4x \: \sqrt{-g} \: (RR)^{\mu\nu\alpha\beta}
(\tilde{\Delta}^{-1}_2 R)_{\mu\nu\alpha\beta} \; .
\end{equation}
It may also be possible to obtain such actions by a descent
procedure from conformal invariants in the regularized,
D=4-2$\epsilon$ dimension.

\section{Summary}

Use of a novel class of tensorial conformal invariant operators
has made possible compact closed form expressions for type B
effective actions in any dimension; these retain all the physical
``UV" characteristics of the underlying matter loop integrals from
which they arise.  By contrast the known type A actions are
(beyond D=2) far from reflecting those origins.  Nevertheless, it
was possible to verify that, when properly reformulated, their
leading terms have only single poles, as required by dimension and
general anomaly considerations and therefore bound to be fulfilled
by any action that yields the correct anomaly. Given the ongoing
popularity of type A, in problems ranging from phonomenological
gravitational actions to C-theorems and holography (for some
recent work see {\it e.g.}, \cite{balbinot,imbimbo} and references
therein), improved versions, perhaps of the suggested form (32),
of its effective action are eminently worth finding.

\noindent{\bf Acknowledgements}

I thank T.\ Branson for discussions leading to his derivation of
the conjectured $\tilde{\Delta}_4$ operator, A.\ Schwimmer for
extremely stimulating exchanges that included keeping the worst
errors out and H.\ Osborn for useful correspondence. This work was
supported by the National Science Foundation under grant
PHY99-73935.

\end{document}